\documentclass[aps,prl,twocolumn,superscriptaddress,showpacs,floatfix]{revtex4-2}
\usepackage{comment}
\usepackage{physics}
\usepackage{amsmath,amsthm,amssymb,amsfonts, color, comment, graphicx}%, %fancyhdr, environ}
\usepackage{xcolor}
\usepackage{float}
\usepackage{graphicx}
\usepackage{dcolumn}% Align table columns on decimal point
\usepackage{bm}% bold math
\UseRawInputEncoding
\usepackage{pgffor}
\usepackage{pdfpages}
\makeatletter
\AtBeginDocument{\let\LS@rot\@undefined}
\makeatother
\usepackage{hyperref}

\usepackage[normalem]{ulem}

\newcommand{\bea}{\begin{eqnarray}}
\newcommand{\eea}{\end{eqnarray}}

\newcommand{\be}{\begin{equation}}
\newcommand{\ee}{\end{equation}}

\newcommand{\beal}{\begin{align}}
\newcommand{\eeal}{\end{align}}

\newcommand{\upa}{\uparrow}
\newcommand{\dna}{\downarrow}

\newcommand{\btjstrw}{\mathrel{{\rotatebox[origin=c]{90}
{$\bowtie$}}\kern-0.18em\raisebox{-.95ex}{$\bullet$}
\kern-0.5em\raisebox{.97ex}{$\bullet$}
\kern-1.12em\raisebox{.97ex}{$\bullet$}
\kern-0.52em\raisebox{-.95ex}{$\bullet$}}}

\newcommand{\btjnbrR}{{\mathrel{\rotatebox[origin=c]{90}
{$\bowtie$}}\kern-0.22em\raisebox{.9ex}{$\bullet$}
\kern-1.em\raisebox{-.8ex}{$\bullet$}}}
\newcommand{\btjnbrL}{{\mathrel{\rotatebox[origin=c]{90}
{$\bowtie$}}\kern-0.22em\raisebox{-.8ex}{$\bullet$}
\kern-1.em\raisebox{+.9ex}{$\bullet$}}}

\def\s{\sigma}
\def\t{\tau}

\def\w{\omega}
\def\x{\xi}

\def\P{\Pi}

\def\W{\Omega}

\newcommand{\llangle}[1][]{\savebox{\@brx}{\(\m@th{#1\langle}\)}%
  \mathopen{\copy\@brx\kern-0.5\wd\@brx\usebox{\@brx}}}
\newcommand{\rrangle}[1][]{\savebox{\@brx}{\(\m@th{#1\rangle}\)}%
  \mathclose{\copy\@brx\kern-0.5\wd\@brx\usebox{\@brx}}}

\begin{document}

\preprint{APS/123-QED}

\title{Enhanced strange metallicity due to Hubbard-$U$ Coulomb repulsion}

\author{Andrew Hardy}
\affiliation{Department of Physics, University of Toronto, 60 St. George Street, Toronto, ON, M5S 1A7 Canada}
\affiliation{Center for Computational Quantum Physics, Flatiron Institute, New York, New York 10010, USA}
\author{Olivier Parcollet}
\affiliation{Center for Computational Quantum Physics, Flatiron Institute, New York, New York 10010, USA}
\affiliation{Universit\'e Paris-Saclay, CNRS, CEA, Institut de Physique Th\'eorique, 91191, Gif-sur-Yvette, France}
\author{Antoine Georges}
\affiliation{Coll\`ege de France, 11 place Marcelin Berthelot, 75005 Paris, France}
\affiliation{Center for Computational Quantum Physics, Flatiron Institute, New York, New York 10010, USA}
\affiliation{CPHT, CNRS, Ecole Polytechnique, Institut Polytechnique de Paris, Route de Saclay, 91128 Palaiseau, France}
\affiliation{DQMP, University of Geneva, 24 quai Ernest-Ansermet, 1211 Geneva, Switzerland}
\author{Aavishkar A. Patel}
\affiliation{Center for Computational Quantum Physics, Flatiron Institute, New York, New York 10010, USA}

\date{\today}

\begin{abstract}
We solve a model of electrons with Hubbard-$U$ Coulomb repulsion and a random
Yukawa coupling to a two-dimensional bosonic bath, using an extended dynamical
mean field theory scheme. Our model exhibits a quantum critical point, at which
the repulsive component of the electron interactions strongly enhances the
effects of the quantum critical bosonic fluctuations on the electrons, leading
to a breakdown of Fermi liquid physics and the formation of a strange metal
with `Planckian' ($\mathcal{O}(k_B T/\hbar)$) quasiparticle decay rates at low
temperatures $T\rightarrow 0$. Furthermore, the eventual Mott transition that
occurs as the repulsion is increased seemingly bounds the maximum decay rate in
the strange metal. Our results provide insight into low-temperature
strange metallicity observed in proximity to a Mott transition, as is observed,
for instance, in recent experiments on certain moir\'{e} materials.     
\end{abstract}

\maketitle

{\it Introduction:}
Understanding the properties of itinerant electron systems proximate to a quantum critical point (QCP) remains a major open problem in condensed matter physics. These systems could possibly describe the enigmatic strange metals, which have linear-in-temperature resistivity at low temperatures $T \rightarrow 0$ \cite{SachdevKeimer2011}. Strange metal behavior has been observed in a wide variety of strongly correlated systems \cite{Hayes2016, Nakajima2020, legros2019, Nguyen2021, jiang2023, jaoui2022, ghiotto2021}, the most iconic of these being proximate to a putative two-dimensional (2D) antiferromagnetic critical point, as seen in Fe- and also possibly CuO- based superconductors \cite{Hayes2016, Nakajima2020, legros2019, Gael21, Scalapino2012, keimer2015}. Since strange metals are often the parent states out of which unconventional pairing emerges, understanding them is a prerequisite for a comprehensive theory of correlated-electron superconductivity. 

The properties of low-temperature strange metals are well described by the ``marginal Fermi liquid" (MFL) phenomenology \cite{Varma1989, Varma2020, michon2023a}, in which a particular scaling form of the electron self energy gives rise to the observed linear-in-energy quasiparticle decay and transport scattering rates. Recently, simple models of 2D metallic QCPs \cite{hertz, millis} with spatially random fermion-boson Yukawa interactions \cite{patelUniversalTheoryStrange2023, aldape2022}, were proposed as a generic mechanism for the MFL phenomenology of strange metals, within a perturbative Eliashberg approach.

The Eliashberg approach utilized in Refs. \cite{patelUniversalTheoryStrange2023, aldape2022} is justifiable only in the limit of a large number of fermion flavors \cite{esterlisLargeTheoryCritical2021}, which eliminates important physical features of realistic SU(2) electrons, such as their Hubbard-$U$ Coulomb repulsion. Recent observations of  quantum criticality and strange metals \cite{ghiotto2021, Zang2022} proximate to Mott transitions in 2D transition metal dichalcogenide moir\'{e} bilayers \cite{tang2020, Wu2018, Lee2023} have underscored the need to incorporate the effects of such repulsion into models of strange metal QCPs. In this Letter, we address this issue using a model of SU(2) electrons with both Hubbard-$U$ Coulomb repulsion and spatially random Yukawa interactions mediated by 2D bosons, that we solve using an extended dynamical mean field theory (EDMFT) method \cite{Sengupta1995, Chitra2000, Smith2000}. 

Using this model, we find that not only do the MFL and strange metal behaviors associated with the QCP survive into non-perturbative regimes, but that they are enhanced by the repulsive interactions. The Hubbard-$U$ reduces the kinetic energy of the electrons, leading to increased local moment formation and an increased  effect of the Yukawa coupling. This leads to `Planckian' quasiparticle decay rates $\sim \alpha_0 k_B T/\hbar$ \cite{HartnollReview}, where $\alpha_0 \sim \mathcal{O}(1)$ and reaches its largest value just before the repulsion-driven Mott transition. 
{\it Model:} 
We consider the following Hamiltonian $H = H_c + H_Y$ that we will use to set up a single-site EDMFT impurity problem:
\begin{equation}
\begin{gathered} 
H_c = -\sum_{ij\s}\left(t_{ij}+\mu \delta_{ij}\right) c_{i\s}^{\dagger} c_{j\s} + U\sum_{i}n_{i\upa}n_{i\dna}, \\
H_Y = \sum_{ir}\frac{{g'}_{ir}}{\sqrt{V}}(-1)^i\vec{S}_i\cdot\vec{\phi}_{r},~~\ll {g'}_{ir} g'_{jr'} \gg = {g'}^2\delta_{ij}\delta_{rr'}.
\label{Hamiltonian}
\end{gathered}
\end{equation}
Here, $i,j,r$ label the sites of a 2D lattice with volume $V$ occupied by the SU(2) electrons and the $3$-component bosonic field $\vec{\phi}$ that represents antiferromagnetic (N\'{e}el) order.
$H_c$ describes a Hubbard model, which is augmented by the Yukawa term $H_Y$ that couples the electron spin $\vec{S}_i$ at each site $i$ to all sites $r$ of the 2D bosonic bath via spatially-random interactions $g'_{ir}$, which are independent random Gaussian variables with zero mean and variance ${g'}^2$. 
The bosonic bath has a conventional imaginary-time ($\t$) Lagrangian
\begin{equation}
\begin{gathered} 
\mathcal{L}_{\vec{\phi}} = \sum_r \frac{1}{2}\left[|\partial_\tau\vec{\phi}_{r\t}|^2 + |\nabla\vec{\phi}_{r\t}|^2 + m_b^2|\vec{\phi}_{r\t}|^2 \right] \\ - i\sum_r\lambda_{r\t}\left[|\vec{\phi}_{r\t}|^2 - 3\kappa\right],
\label{Lagrangian}
\end{gathered}
\end{equation}
which contains the standard kinetic and mass terms, as well as a boson self-interaction that is expressed in the non-linear sigma model fashion using the Lagrange multiplier field $\lambda$ \cite{SachdevQPTbook}. 

In the thermodynamic limit of large volume $V$, the model is controlled by a saddle point that is obtained by averaging over the random coupling $g'_{ir}$ using replicas (Supplementary Information), as in the Sachdev-Ye-Kitaev models \cite{Sachdev2015Bekenstein, Chowdhury2021}. This introduces the bilocal fields $D(\t-\t') = (1/(3V))\sum_r\langle\vec{\phi}_{r\t}\cdot\vec{\phi}_{r\t'}\rangle$ and $\Pi(\t-\t') = ({g'}^2/(3V))\sum_i \langle \vec{S}_{i\t} \cdot \vec{S}_{i\t'}\rangle$. 
Here $D(\t-\t')$ is the local ({\it i.e.} spatially averaged) propagator of the bosonic bath, and $\Pi(\t-\t')$ is the Lagrange multiplier field that enforces its definition and functions as the spatially-uniform self-energy of the bosons $\vec{\phi}$. 
The bilocal field $D(\t-\t')$ mediates a dynamical local spin-spin interaction for the electrons, which enters into the effective local impurity action $S_\mathrm{imp}$ of the EDMFT scheme \cite{Sengupta1995, Chitra2000, Smith2000}:  
\begin{eqnarray}
&&S_{\mathrm{imp}} = \int d\t\left[\sum_\s c_{\s\t}^{\dagger}\left(\partial_\t-\mu\right) c_{\s\t}+U n_{\upa\t} n_{\dna\t}\right] + \\
&&\int d\t d\t' \left[\sum_\s \Delta\left(\t-\t'\right) c_{\s\t}^{\dagger}c_{\s\t'} - \frac{{g'}^2 D\left(\t-\t'\right)}{2} \Vec{S}_{\t} \cdot \Vec{S}_{\t'}\right]. \nonumber
\end{eqnarray}
The hybridization $\Delta(\t -\t')$ maps the local problem to the full dispersing lattice \cite{Georges1996}; we use the hybridization function $\Delta(\t-\t') = t^2 G(\t-\t')$, which is a convenient choice for metallic systems with local propagator $G$ \footnote{This hybridization function is formally exact only for a Bethe lattice, or for all-to-all and random $t_{ij}$ \cite{Georges1996}. However, since we are considering metallic systems with a finite density of states, it should continue to produce the correct qualitative behavior even for a 2D system.}, and is computed self-consistently \cite{dumitrescuPlanckianMetalDopinginduced2022}.
The dynamical interaction $D(\t-\t')$ is also determined self-consistently by solving the bosonic problem with effective action $S_{\mathrm{bath}} = \int d\t \mathcal{L}_{\vec{\phi}} - (1/2)\int d\t d\t' \Pi(\t-\t') \sum_r \vec{\phi}_{r\t}\cdot\vec{\phi}_{r\t'}$, with $\Pi(\t-\t') = ({g'}^2/3)\langle \vec{S}_\t\cdot\vec{S}_{\t'}\rangle$ determined from the impurity problem.
Finally, the electron self energy $\Sigma$ is given by $\Sigma(i\omega_n) = i\omega_n + \mu - \Delta(i\omega_n) - G^{-1}(i\omega_n)$, where $\omega_n \equiv (2n + 1)\pi T$ are fermionic Matsubara frequencies. Unless otherwise specified, we will work with $t=1$, and non-perturbative interaction strengths $U=2t$, ${g'}^2=2t$. The impurity problem is solved using a continuous time Quantum Monte Carlo based on a double expansion in hybridization and spin-spin interaction (CT-SEG) \cite{WernerCTSEG, otsuki2013, gullContinuoustimeMonteCarlo2011}.

We solve the bosonic problem by taking the saddle point of $S_\mathrm{bath}$ with respect to $\lambda$ \footnote{This approximation is valid for 3-component bosons in 2D at or above their upper critical dimension, requiring a dynamical critical exponent $z\geq 2$ \cite{SachdevQPTbook}, which we shall subsequently show is the case here}. 
We also approximate the lattice boson dispersion using a quadratic dispersion and a cutoff $\Lambda_b$, which does not affect the low-energy properties that we will be interested in. This yields the convenient analytic expression
\begin{equation}
D(i\W_n) = \frac{1}{4 \pi} \ln(\frac{\tilde{m}^2_b + \W_n^2 - \Pi(i\W_n) + \Lambda_b^2}{\tilde{m}^2_b + \W_n^2 - \P(i\W_n)}), 
\end{equation}
with bosonic Matsubara frequencies $\Omega_n \equiv 2n\pi T$, and the constraint $D(\t=0)=\kappa$ that is satisfied by adjusting $\tilde{m}_b^2\equiv m_b^2-i\lambda$.

\begin{figure}
\centering
\includegraphics[width=0.48\textwidth]{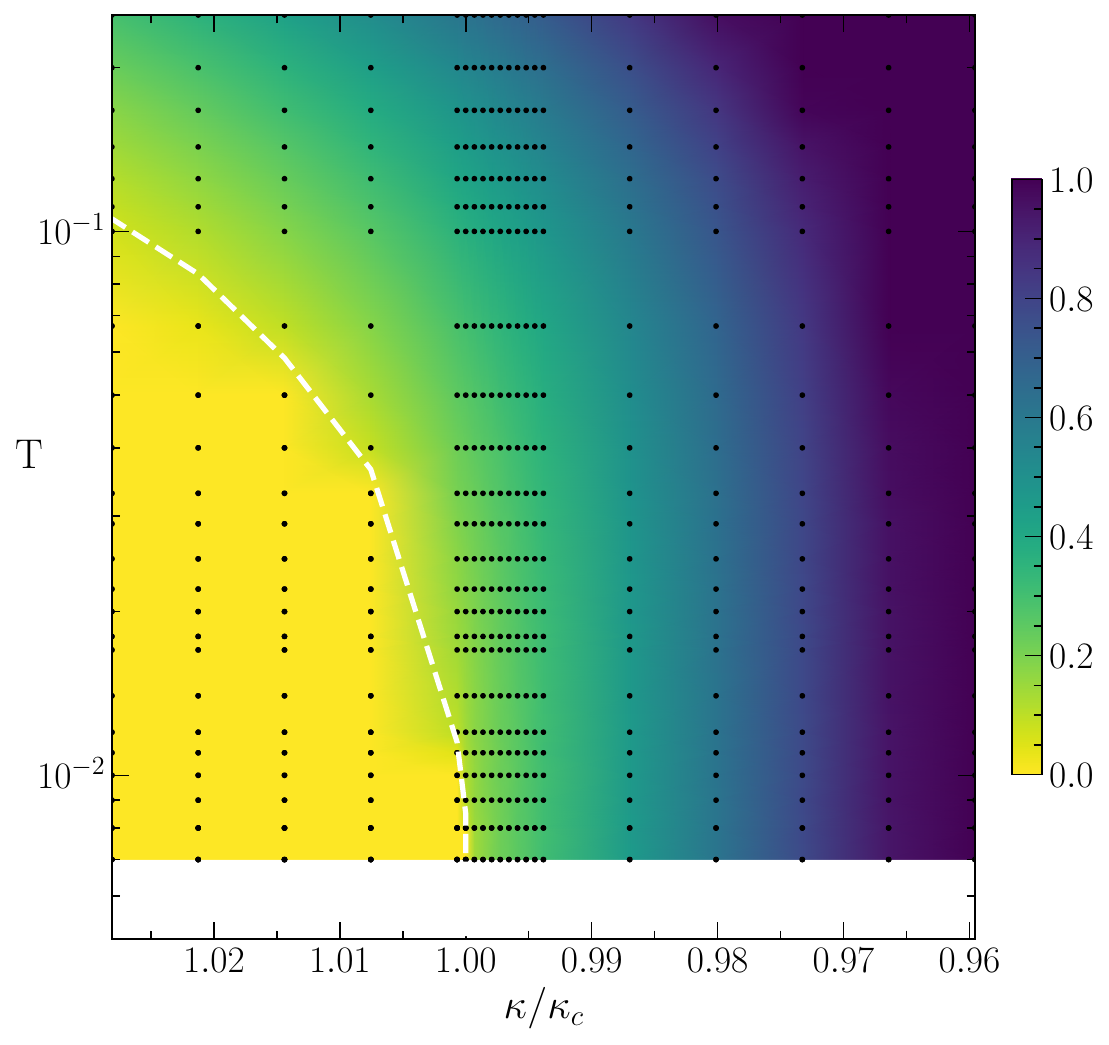}
\caption{Plot of the effective boson mass $m^2 = m^2_b - \Pi(0)$ as a function of the tuning parameter $\kappa$. The dashed line indicates a ``phase boundary" beyond which the bosons develop quasi-long range order and $m^2$ becomes small. The phase boundary extrapolates to a QCP ($\kappa = \kappa_c$) as $T \rightarrow 0$.}
\label{phasediagram}
\end{figure}

In order to probe strange metal physics, we need to tune the system to a QCP at which the bosons are gapless and coupling to them can destroy the electron quasiparticles \cite{SachdevQPMbook}. The value of $\kappa$ controls the value of $\tilde{m}_b^2$, and therefore the value of the boson gap $m^2 = \tilde{m}_b^2 - \Pi(0)$. The QCP can then be identified as a critical $\kappa = \kappa_c$, for which $m^2(T \rightarrow 0, \kappa \leq \kappa_c) = 0$ but $m^2(T \rightarrow 0, \kappa > \kappa_c) \neq 0$. Fig. \ref{phasediagram} shows our numerical result for $m^2$ as a function of $\kappa$ and $T$, which establishes the existence of such a QCP in the bosonic sector. 
The frequency dependent part of $\Pi(i\Omega_n)$ is $\sim |\Omega_n|^{y}$ with $y < 1$ at low frequencies (Supplementary Information), which implies a dynamical critical exponent of $z > 2$ for the quadratically dispersing 2D bosons.

{\it Marginal Fermi liquid:} We determine the fate the electron quasiparticles upon approach to such a QCP by analyzing the electron self-energy $\Sigma(i\omega_n)$. 
In a Fermi liquid (FL) with well-defined quasiparticles with quasiparticle residue $Z$, $-\mathrm{Im}[\Sigma(i\omega_n)] \approx \omega_n (Z^{-1} - 1) + A(\pi^2 T^2 - \omega_n^2)\mathrm{sgn}(\omega_n)$, leading to a quasiparticle decay rate that scales quadratically in $\omega,~T$ upon analytic continuation to real frequencies $\omega$. 
At the first Matsubara frequency $\omega_1 = \pi T$, $-\mathrm{Im}[\Sigma(i\omega_1)] \approx \pi T (Z^{-1} - 1)$ \cite{Maslov2012a}. 
In contrast, the breakdown of FL physics and the destruction of quasiparticles at low energies is characterized by a non-Fermi liquid behavior of $-\mathrm{Im}[\Sigma(\Lambda_f \gg |i\omega_n| \gg T)] \sim |\omega_n|^x\mathrm{sgn}(\omega_n)$, with $x < 1$, where $\Lambda_f$ is an ultraviolet cutoff.
Additionally, at the first Matsubara frequency $\omega_1 = \pi T$, $-\mathrm{Im}[\Sigma(i\omega_1)] \sim T^x$ \cite{SachdevQPMbook}. The special case of $x \rightarrow 1^-$ is termed a MFL \cite{Varma1989}, where $-\mathrm{Im}[\Sigma(\Lambda_f \gg |i\omega_n| \gg T)] \sim \omega_n \ln(\Lambda_f/|\omega_n|)$ and $-\mathrm{Im}[\Sigma(i\omega_1)] \sim T \ln(\Lambda_f/T)$. For real frequencies, MFL behavior leads to the linear-in-$\omega$ and linear-in-$T$ quasiparticle decay rates observed experimentally in strange metals \cite{michon2023a}. 

\begin{figure}[H]
     \centering
     \includegraphics[width=0.48\textwidth]{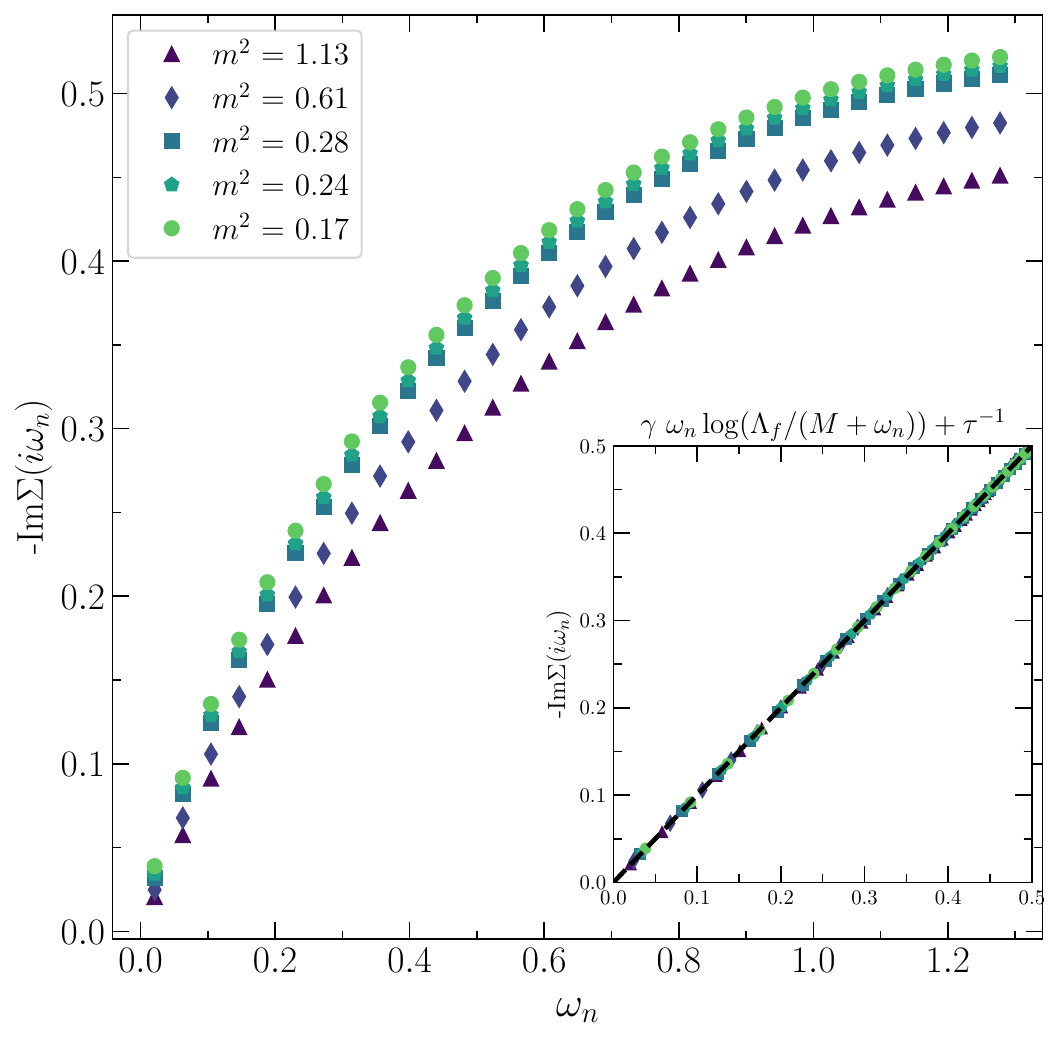}
     \caption{The self-energy -$\operatorname{Im}[\Sigma(i \w_n)]$ upon approaching $\kappa=\kappa_c$ at $T = 0.00\overline{6}$. Lighter colors indicate increasing $\kappa$ and  correspondingly, a decreasing $m^2$. The inset shows the fit to Eq. \ref{functionalform}. The points lie on the $y = x$ line, indicating an excellent fit.}
    \label{self_energy}
\end{figure}

We expect that $\Sigma(i\omega_n)$ will cross over from FL to MFL behavior as the QCP is approached by tuning $\kappa$. We therefore consider a `universal' scaling ansatz 
\begin{equation}
    -\operatorname{Im}[\Sigma(i\omega_n)] = \gamma  \omega_n\ln(\frac{\Lambda_f}{M +  |\omega_n|}) + \tau^{-1} \mathrm{sgn}(\omega_n);
    \label{functionalform}
\end{equation}
here the MFL scaling is modified by an infrared cutoff $M$, which accounts for the presence of a finite boson gap and which should vanish as $T \rightarrow 0$ at the QCP. Away from the QCP, $M>0$, and $-\mathrm{Im}[\Sigma(i\omega_n)]$ is an analytic function of $\omega_n$, as in a FL. The quasiparticle decay rate arising from thermal fluctuations, which should vanish at $T=0$, is given by $-\text{Im}[\Sigma(0^+)] = \tau^{-1}$. The effective coupling $\gamma$, as well as $\Lambda_f$, should both be independent of $m^2$. 
This scaling form is motivated by the perturbative calculation of $\Sigma(i\omega_n)$ (Supplementary Information). 
As shown in Fig. \ref{self_energy}, we obtain an excellent fit to Eq. \ref{functionalform} for sizeable $g',~U$, thereby establishing MFL behavior in a non-perturbative regime. 
\begin{figure*}
         \centering
         \includegraphics[width=\textwidth]{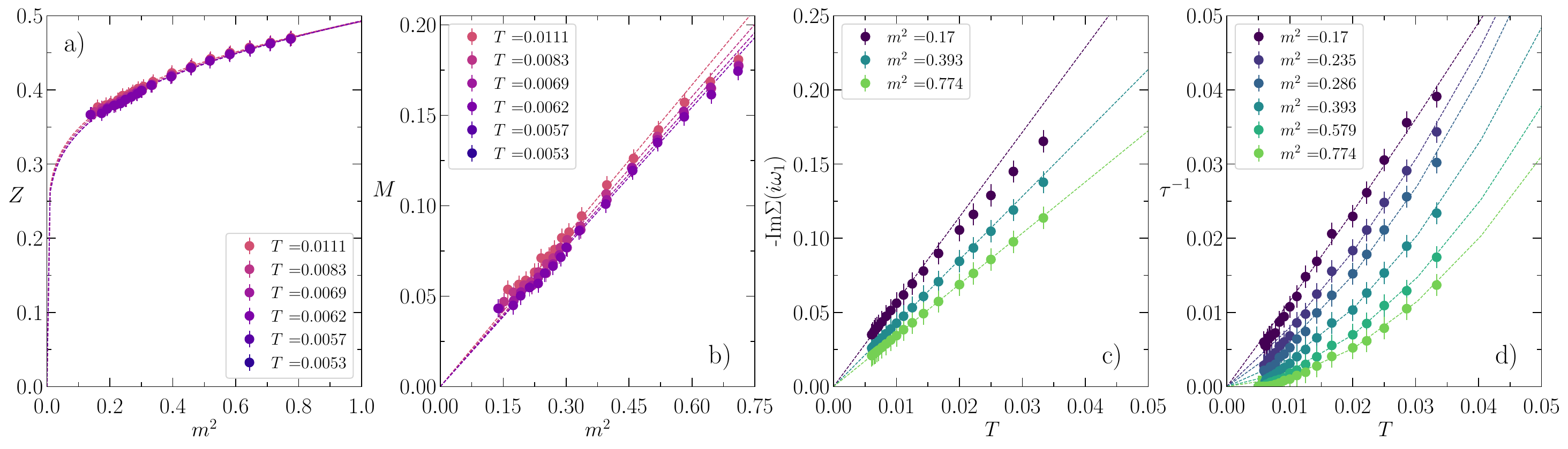}
         \caption{a) Plot of the fermion quasiparticle residue $Z$ as a function of the effective boson mass $m^2 = \tilde{m}^2_b - \Pi(0)$. The fit lines are of the form $Z = (1+ a_1 \log( a_2/m^2))^{-1}$ b) The IR cut-off $M$ in  Eq.\ref{functionalform}, which is linearly proportional to $m^2$. The fit lines are linear fits. c) $-\Im[\Sigma(\omega_1)]$ as a function of $T$, showing a deviation from the expected linear FL behavior upon approaching the QCP (reducing $m^2$), with the temperature scale below which $-\mathrm{Im}[\Sigma(i\omega_1)] \propto T$ reducing as the QCP is approached. The fit lines are linear fits to the low $T$ data points. d) The quasiparticle decay rate $\tau^{-1}(T)$, which transitions from a $T^2$ to a linear-in-$T$ dependency upon approaching the QCP (reducing $m^2$). The fit lines are of the form $\tau^{-1} = \alpha_0 T + \alpha_1 T^2$. The values of $m^2$ shown in panels c) and d) are at the lowest temperatures shown.}
        \label{criticality}
\end{figure*}

The behavior of various parameters associated with Eq. \ref{functionalform} is shown in Fig. \ref{criticality}. Fig. \ref{criticality}a) shows the quasiparticle residue $Z$, computed through the extracted parameters as $Z = \left(1-\partial_{\omega_n} \Im(\Sigma(i\omega_n))\right)^{-1} |_{\omega_n \rightarrow 0^+} = \left(1+ \gamma\log(\Lambda_f/M)\right)^{-1}$. 
The decrease of $Z$ as $m^2$ is reduced indicates the destruction of quasiparticles as the QCP is approached. Fig. \ref{criticality}b) shows that the IR cutoff $M$ decreases linearly with $m^2$, extrapolating to $M \rightarrow 0$ as $m^2 \rightarrow 0$, thereby establishing quantum critical behavior in the electron sector as well. 
An independent check that does not involve Eq. \ref{functionalform} is shown in Fig. \ref{criticality}c), where we observe a clear deviation from the FL scaling of $-\Im[\Sigma(i \omega_1,T)] \propto T$ as $m^2$ is reduced and the QCP is approached. 
The behavior of $\gamma$ and $\Lambda_f$ is provided in the Supplementary Information, where we find that they are rather insensitive to $m^2$.

The quasiparticle decay rate $\tau^{-1}$ is shown in Fig.\ \ref{criticality}d). The $T$-dependence of $\tau^{-1}$ is fit well by $\tau^{-1}(T) = \alpha_0 T + \alpha_1 T^2$. As $m^2$ is decreased and the QCP is approached, The coefficient of the $T$-linear term $\alpha_0$ grows by orders of magnitude, and $\tau^{-1}(T)$ crosses over to a linear scaling with $T$ with $\alpha_0 \sim \mathcal{O}(1)$, which is reminiscent of the `Planckian' decay rates observed in experiments on strange metals \cite{HartnollReview}. In the single-site EDMFT scheme used here, for a Fermi surface with a large Fermi energy and a finite density of states, the transport scattering rate is also determined by $\tau^{-1}$ \cite{Georges1996}, leading to strange metallic $T$-linear resistivity at the QCP.

\begin{figure}
 \centering
 \includegraphics[width=0.48\textwidth]{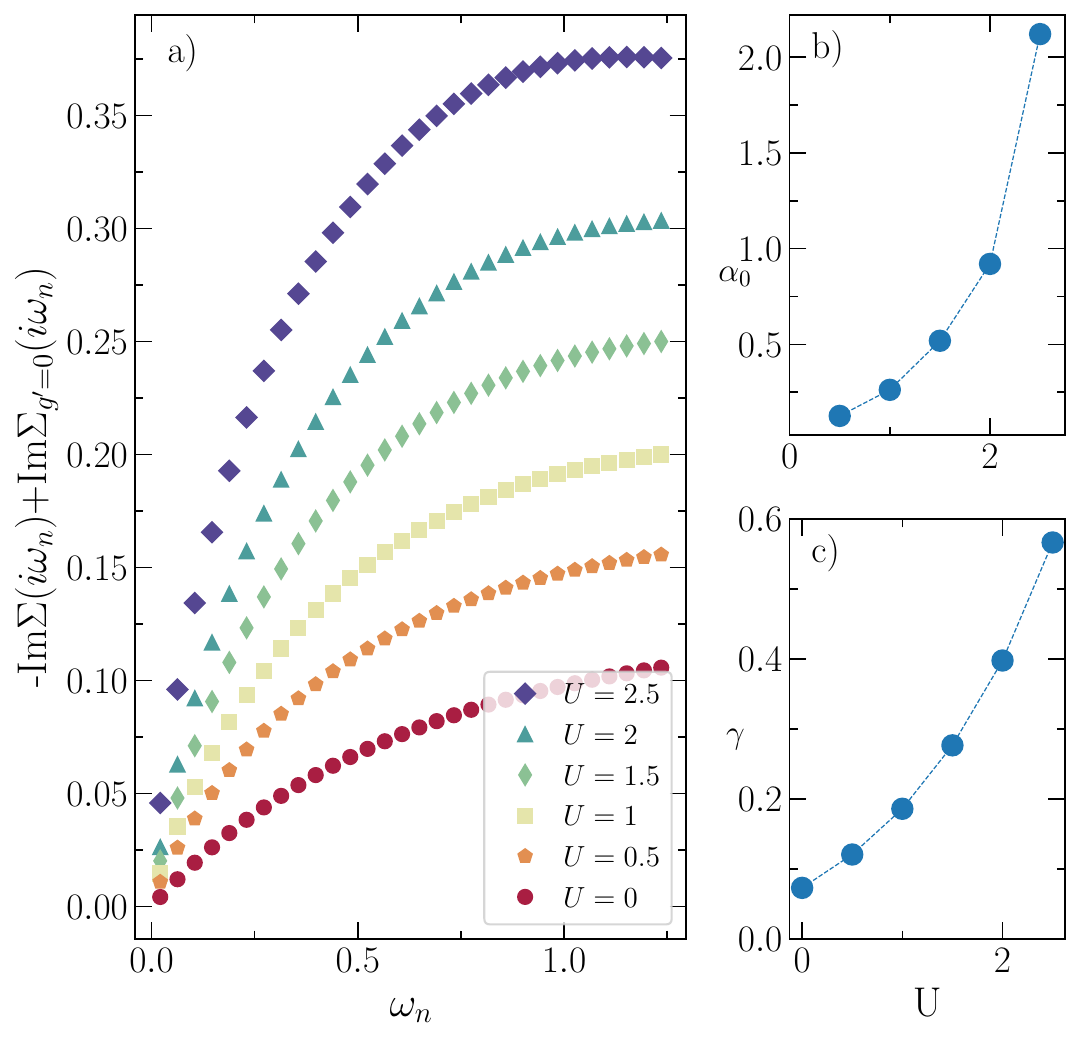}
 \caption{a) The contribution of the critical bosonic fluctuations to $-\mathrm{Im}[\Sigma(i\omega_n)]$ for different values of $U$ and fixed $g'$ at $T = 0.00\overline{6}$. b) The slope $\alpha_0$ of the $T$-linear quasiparticle decay rate as a function of $U$. c) The coefficient $\gamma$ of the fit to Eq. \ref{functionalform} as a function of $U$, which determines the strength of the MFL.}
\label{U_enhancement}
\end{figure}

We finally turn to the main result of this Letter. We consider $\Sigma(i \omega_n)$ for systems tuned to criticality with increasing values of $U$. Fig. \ref{U_enhancement} a) shows $-\mathrm{Im}[\Sigma(i\omega_n)]$ with its value at $g' = 0$ subtracted off, thereby isolating the contribution of the critical bosonic fluctuations to the self energy. This contribution increases significantly with increasing $U$, leading to a strong enhancement of strange metal and MFL behavior as shown in Fig. \ref{U_enhancement}b, c. The slope $\alpha_0$ of the $T$-linear quasiparticle decay rate $\tau^{-1}(T) \approx \alpha_0 T$ increases rapidly as $U$ is increased, as does the coefficient $\gamma$ of the MFL frequency dependence in Eq. \ref{functionalform}. 

Increasing $U$ beyond $U \approx 2.5t$ (at ${g'}^2=2t$) leads to a Mott transition before criticality can be reached at low $T$ (Supplementary Information), which seemingly establishes a bound on the maximum value of $\alpha_0$. This is reminiscent of the ``stability bounds" on $T$-linear decay rates noted empirically in models of systems with well-defined quasiparticles and attractive electron-phonon interactions \cite{Murthy2023}, but arises here in a different context of quantum critical systems with poorly defined quasiparticles and with a combination of attractive (Yukawa) and repulsive (Hubbard) interactions. 

{\it Conclusion and Outlook:} We have solved a model of electrons with a spatially random Yukawa coupling to a 2D bosonic bath. We find a quantum critical point that gives rise to marginal Fermi liquid and strange metal behaviors. These behaviors are strongly enhanced by Hubbard-$U$ repulsive interactions between the electrons, which allows for large $T$-linear quasiparticle decay rates in the strange metal that are around the `Planckian' value of $k_B T/\hbar$, with the largest values occurring just before the Mott transition. This amplification of quantum critical random Yukawa interactions by Coulomb repulsion might help explain the `Planckian' decay rates in strange metals recently observed near Mott transitions in certain moir\'{e} materials \cite{ghiotto2021}, and also indicates that a relatively modest bare amount of randomness in electron interactions might be sufficient to obtain strange metal behavior. Our results may also indicate why strange metals that are not in proximity to a conduction-band Mott transition demonstrate sub-Planckian coefficents in their scattering rate ($\alpha_0 < \mathcal{O}(1)$) \cite{Taupin2022}.

Several interesting directions exist for future studies: recent work \cite{Li2024} has studied superconductivity arising from random Yukawa interactions in an Eliashberg approach without repulsion. It would be interesting to examine if the repulsion-enhanced quantum critical bosonic fluctuations described in this Letter can lead to an amplification of superconductivity that exceeds any suppression by the bare repulsion itself. It would also be interesting to investigate if models like the ones considered here can produce crossovers to `bad metals' with $\mathcal{O}(\hbar/e^2)$ resistivities at high temperatures, and to compute the transport properties of such regimes \cite{Deng2013}.

Random Yukawa models have also been studied recently without assumptions of disorder self-averaging \cite{patel2024, AAPQMC}; it was found that the quantum critical point is replaced by a gapless boson phase that also gives rise to strange metal behavior. It would be interesting to explore the effects of the Hubbard-$U$ repulsion in such a scenario using multi-site EDMFT methods. Finally, the dynamical cluster approximation \cite{Hettler2000}, which restores momentum-space structure and thereby a finite average antiferromagnetic coupling in the effective impurity problem, can be applied to our problem to explore the interplay between pseudogap and strange metal phases.  

\begin{acknowledgments} 
\textit{Acknowledgments:} We thank A. Hampel, N. Kavokine, F. Kugler, S. Sachdev and N. Wentzell for helpful comments. A. H. acknowledges support from a NSERC Graduate Fellowship (PGS-D) and a predoctoral fellowship at the Flatiron Institute while this work was being completed. The algorithms used in this study were implemented using the TRIQS library  \cite{Kavokine_JOSS_CTSEG, kavokine2024, parcolletTRIQSToolboxResearch2015}. 
The Flatiron Institute is a division of the Simons Foundation.
\end{acknowledgments}

\bibliography{apssamp}% Produces the bibliography via BibTeX.

\newpage


\section*{Supplementary material}

# Supplementary Information for
# Enhanced strange metallicity due to Coulomb repulsion

Andrew Hardy,[1, 2] Olivier Parcollet,[2, 3] Antoine Georges,[4, 2, 5, 6] and Aavishkar A. Patel[2]

[1]*Department of Physics, University of Toronto, 60 St. George Street, Toronto, ON, M5S 1A7 Canada*

[2]*Center for Computational Quantum Physics, Flatiron Institute, New York, New York 10010, USA*

[3]*Université Paris-Saclay, CNRS, CEA, Institut de Physique Théorique, 91191, Gif-sur-Yvette, France*

[4]*Collège de France, 11 place Marcelin Berthelot, 75005 Paris, France*

[5]*CPHT, CNRS, Ecole Polytechnique, Institut Polytechnique de Paris, Route de Saclay, 91128 Palaiseau, France*

[6]*DQMP, University of Geneva, 24 quai Ernest-Ansermet, 1211 Geneva, Switzerland*

(Dated: August 28, 2024)

## I. SADDLE POINT AND EDMFT APPROXIMATION

We average the action $S = \int d\tau (\sum_{i\sigma} c^\dagger_{i\sigma\tau} \partial_\tau c_{i\sigma\tau} + H + \mathcal{L}_{\vec{\phi}})$ over realizations of the disordered Yukawa couplings $g'_{ir}$, to obtain an effective action $S_{\text{eff}}$. The effective action is defined using the disorder-averaged partition function $\mathcal{Z} = \int \mathcal{D}[g'_{ir}] \mathcal{P}(g'_{ir}) e^{-S} / \int \mathcal{D}[g'_{ir}] \mathcal{P}(g'_{ir}) \equiv e^{-S_{\text{eff}}}$, where $\mathcal{P}(g'_{ir}) = e^{-\sum_{ir} g'^2_{ir}/(2g'^2)}$. Formally, this process is carried out using replicas [1]. We obtain

$$S_{\text{eff}} = \int d\tau \left[ \sum_{i\sigma} c^\dagger_{i\sigma\tau} \partial_\tau c_{i\sigma\tau} + H_c + \mathcal{L}_{\vec{\phi}} \right] - \frac{g'^2}{2V} \int d\tau d\tau' \sum_{ir} \sum_{ab=1}^{3} \phi^a_{r\tau} \phi^b_{r\tau'} S^a_{i\tau} S^b_{i\tau'}. \tag{1.1}$$

We then introduce the bilocal field $D^{ab}_{\tau\tau'}$, and the Lagrange multiplier bilocal field $\Pi^{ab}_{\tau\tau'}$ that enforces the definition of $D^{ab}_{\tau\tau'}$:

$$\begin{aligned} S_{\text{eff}} = &\int d\tau \left[ \sum_{i\sigma} c^\dagger_{i\sigma\tau} \partial_\tau c_{i\sigma\tau} + H_c \right] - \frac{g'^2}{2} \int d\tau d\tau' \sum_i \sum_{ab=1}^{3} D^{ab}_{\tau\tau'} S^b_{i\tau'} S^a_{i\tau} \\ &+ \frac{V}{2} \int d\tau d\tau' \sum_{ab=1}^{3} \Pi^{ba}_{\tau'\tau} \left( D^{ab}_{\tau\tau'} - \frac{1}{V} \sum_r \phi^a_{r\tau} \phi^b_{r\tau'} \right) + \int d\tau \mathcal{L}_{\vec{\phi}}. \end{aligned} \tag{1.2}$$

In the thermodynamic limit of large system volume $V$, $S_{\text{eff}}$ is dominated by its saddle point with respect to these bilocal fields. The dominance of such a saddle point also ensures that the system self-averages over disorder in this limit. Because of spin-rotation invariance, the components of the bilocal fields that are off-diagonal in $a, b$ vanish at the saddle point. We then get $\Pi^{ab}_{\tau\tau'} = \Pi(\tau - \tau')\delta_{ab}$, and $D^{ab}_{\tau\tau'} = D(\tau - \tau')\delta_{ab}$ at the saddle point, where $\Pi(\tau - \tau')$ and $D(\tau - \tau')$ are defined in the main text. With these saddle point values of the bilocal fields inserted, the first line of Eq. 1.2 is a 2D fermionic problem with spatially local interactions, that becomes the impurity problem $S_{\text{imp}}$ of the main text in the EDMFT approximation [2–4]. The second line is the 2D bosonic problem described by $S_{\text{bath}}$ of the main text. The two problems must be solved together self-consistently.

We can also formulate the problem using infinite-dimensional electrons, which makes the EDMFT approximation exact [5]. To this end, we define the infinite-dimensional lattice coordinate $i \equiv (i_x, i_y, i_z)$, where $i_z = (i_3, i_4, ...)$,

and the 2D lattice coordinate $r \equiv (r_x, r_y)$. We then modify the Hamiltonian to $H_\infty = H_c^\infty + H_Y^\infty$, with

$$H_c^\infty = -\sum_{ij\sigma}(t_{ij} + \mu\delta_{ij})c_{i\sigma}^\dagger c_{j\sigma} + U\sum_i n_{i\uparrow}n_{i\downarrow},$$

$$H_Y^\infty = \sum_{ir}\frac{g'_{ir}}{\sqrt{V}}(-1)^i \vec{S}_i \cdot \vec{\phi}_{i_z r}, \quad \ll g'_{ir}g'_{jr'} \gg = g'^2 \delta_{ij}\delta_{rr'}. \tag{1.3}$$

Hence, each 2D slice (which is indexed by $i_z$ and is of volume $V$) of the infinite-dimensional electron lattice is coupled to a corresponding 2D bosonic bath $\vec{\phi}_{i_z r}$ of volume $V$. The Lagrangian of the stack of 2D baths is given by

$$\mathcal{L}_{\vec{\phi}}^\infty = \sum_{i_z r}\frac{1}{2}\left[|\partial_\tau \vec{\phi}_{i_z r\tau}|^2 + |\nabla_r \vec{\phi}_{i_z r\tau}|^2 + m_b^2|\vec{\phi}_{i_z r\tau}|^2\right] - i\sum_{i_z r}\lambda_{i_z r\tau}\left[|\vec{\phi}_{i_z r\tau}|^2 - 3\kappa\right]. \tag{1.4}$$

Upon averaging over $g'_{ir}$, we obtain the equivalent of Eq. 1.2:

$$S_{\text{eff}}^\infty = \int d\tau\left[\sum_{i\sigma}c_{i\sigma\tau}^\dagger \partial_\tau c_{i\sigma\tau} + H_c^\infty\right] - \frac{g'^2}{2}\int d\tau d\tau' \sum_i \sum_{ab=1}^3 D_{\tau\tau'}^{ab}(i_z)S_{i\tau'}^b S_{i\tau}^a$$
$$+\frac{V}{2}\int d\tau d\tau' \sum_{i_z}\sum_{ab=1}^3 \Pi_{\tau'\tau}^{ba}(i_z)\left(D_{\tau\tau'}^{ab}(i_z) - \frac{1}{V}\sum_r \phi_{i_z r\tau}^a \phi_{i_z r\tau'}^b\right) + \int d\tau \mathcal{L}_{\vec{\phi}}. \tag{1.5}$$

At the large $V$ saddle point, the system obeys translational invariance with respect to $i_z$ as well as spin-rotation invariance, so $\Pi_{\tau\tau'}^{ab}(i_z) = \Pi(\tau - \tau')\delta_{ab}$, and $D_{\tau\tau'}^{ab}(i_z) = D(\tau - \tau')\delta_{ab}$. We then have an infinite-dimensional electron problem with local effective interactions, whose exact solution is given by the results of the main text.

## II. FREQUENCY DEPENDENCE OF $\Pi(i\Omega_n)$

The frequency dependence of $\Pi(i\Omega_n)$ is shown in Fig. 1. At low frequencies, and small values of $m^2$ that correspond to low temperatures at $\kappa = \kappa_c$, $\Pi(0) - \Pi(i\Omega_n)$ has a visibly sub-linear dependence on $|\Omega_n|$, which implies a dynamical critical exponent $z > 2$ for quadratically dispersing bosons.

## III. PERTURBATIVE SOLUTION FOR THE SELF ENERGY

For the non-interacting system, the self-consistent solution for $G(i\omega_n)$ in presence of the hybridization $\Delta(i\omega_n)$ is $G_0 \approx -(i/t)\text{sgn}(\omega_n)$ at low frequencies and $\mu = 0$. This leads to the perturbative boson self energy

$$\Pi_0(i\Omega_n = 0, T = 0) - \Pi_0(i\Omega_n, T = 0) = \frac{g'^2}{2}\int\frac{d\omega_n}{2\pi}\left[G_0\left(i\omega_n + \frac{i\Omega_n}{2}\right)G_0\left(i\omega_n - \frac{i\Omega_n}{2}\right) - G_0^2(i\omega_n)\right]$$
$$\approx \frac{g'^2}{4\pi t^2}|\Omega_n| \equiv c_d|\Omega_n|. \tag{3.1}$$

The perturbative fermion self energy due to $g'$ is given by

$$\Sigma_0^{g'}(i\omega_n, T = 0) = \frac{3g'^2}{4}\int\frac{d\Omega_n}{2\pi}G_0(i\omega_n + i\Omega_n)D_0(i\Omega_n), \tag{3.2}$$

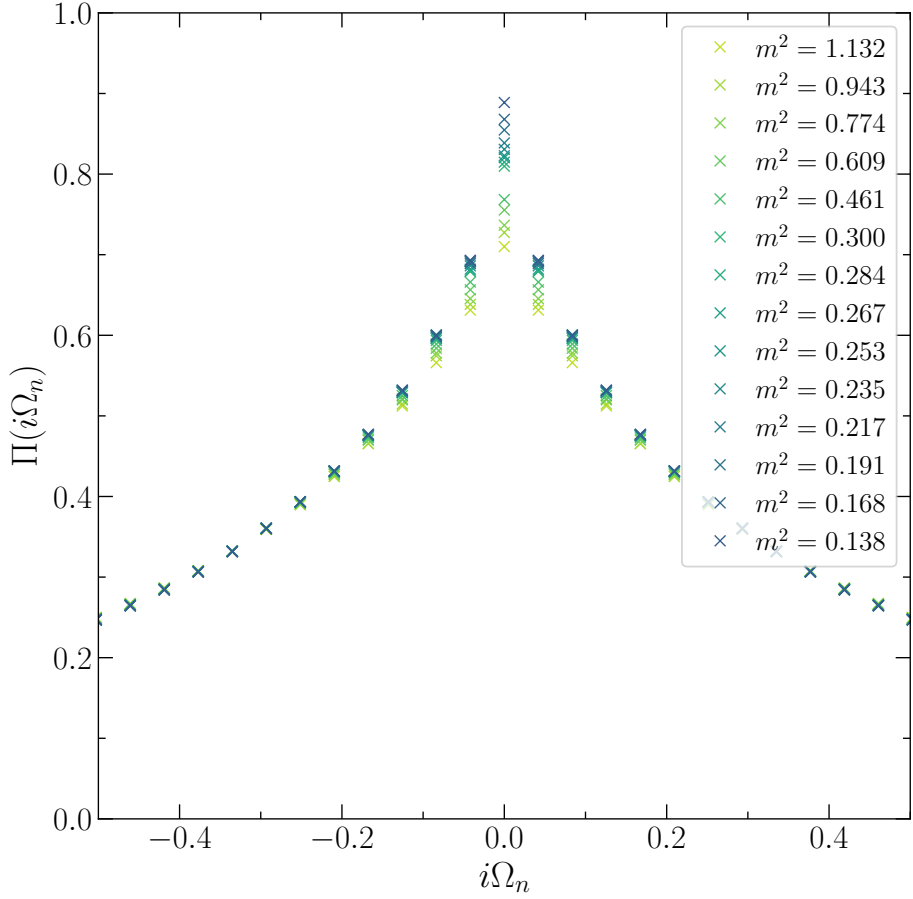

FIG. 1: The frequency dependent boson damping (or self-energy) term $\Pi(i\Omega_n)$.

where $\Pi_0(i\Omega_n)$ is included in the boson Green's function $D_0$ in the perturbative Eliashberg scheme. We then have

$$\begin{aligned}
\Sigma_0^{g'}(i\omega_n, T=0) &= -\frac{3ig'^2}{4t} \int_0^{\Lambda_b} \frac{qdq}{2\pi} \int \frac{d\Omega_n}{2\pi} \frac{\text{sgn}(\omega_n + \Omega_n)}{\Omega_n^2 + q^2 + c_d|\Omega_n| + m^2} \\
&\approx -\frac{3ig'^2}{16\pi t} \int_{-\omega_n}^{\omega_n} \frac{d\Omega_n}{2\pi} \ln\left(\frac{\Lambda_b^2}{c_d|\Omega_n| + m^2}\right) \\
&= -\frac{3ig'^2 m^2}{16 c_d \pi^2 t} \left( \ln\left[\frac{1}{1 + c_d \frac{|\omega_n|}{m^2}}\right] \text{sgn}(\omega_n) + \left(c_d \frac{\omega_n}{m^2}\right) \ln\left[\frac{e\Lambda_b^2/m^2}{1 + c_d \frac{|\omega_n|}{m^2}}\right] \right) \\
&\approx -i\gamma \omega_n \ln\left(\frac{\Lambda_f}{M + |\omega_n|}\right), \quad \omega_n \to 0.
\end{aligned} \quad (3.3)$$

In the limit of small $\omega_n > 0$, Eq. 3.3 is of the form of the universal scaling ansatz of the main text, upon addition of the constant $\tau^{-1}$ to account for thermal fluctuations at $T > 0$. A small value of $U$ simply introduces a perturbative Fermi liquid self energy that is $\Sigma_0^U(i\omega_n) \propto -i\omega_n$ to leading order in $\omega_n$, along with a Hartree shift of $\mu$ that is eliminated by fixing the fermion density. Adding $\Sigma_0^U$ to $\Sigma_0^{g'}$ then simply shifts the value of $\Lambda_f$, preserving the universal scaling ansatz for the total fermion self energy. The fermion self energy also does not affect $G(i\omega_n) \approx G_0(i\omega_n)$ to leading order in $\omega_n$, implying that the above perturbative Eliashberg solution is self-consistent at small couplings.

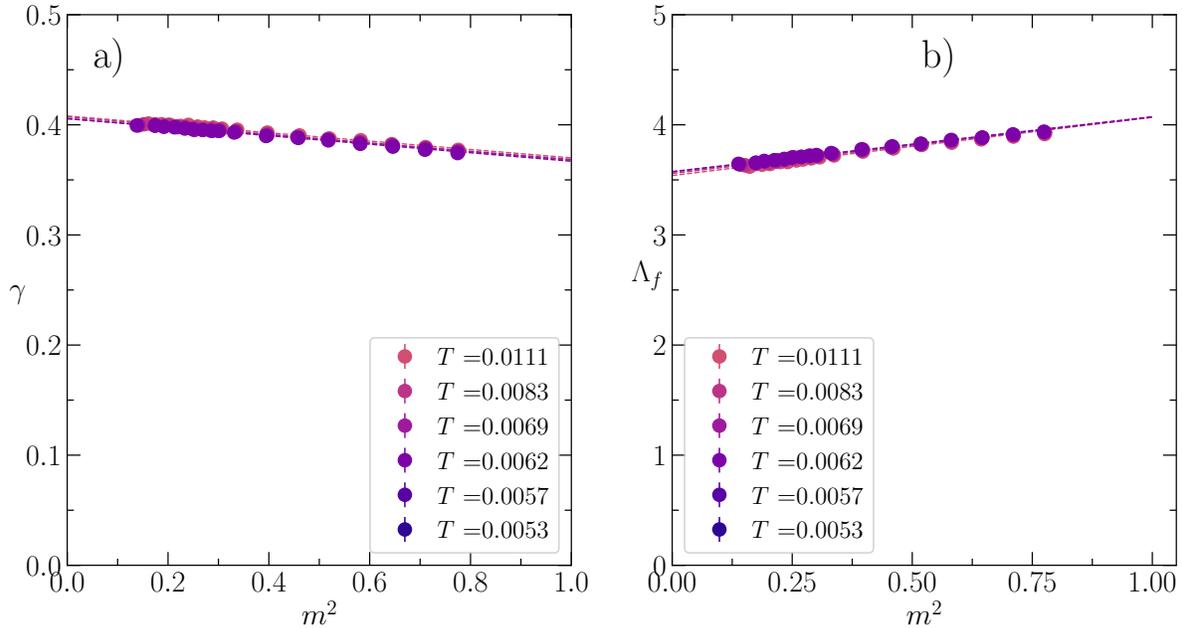

FIG. 2: The parameters $\gamma$ and $\Lambda_f$, of the fit to the universal scaling ansatz for $-\mathrm{Im}[\Sigma(i\omega_n)]$, which are more or less independent of $T$ and $m^2$.

## IV. FIT PARAMETERS

We demonstrate in Fig. 2 that the fit parameters $\gamma$ and $\Lambda_f$ of the universal scaling ansatz for $-\mathrm{Im}[\Sigma(i\omega_n)]$ are independent of $T$ and only depend very weakly upon $m^2$.

## V. MOTT TRANSITION

As mentioned in the main text, the critical point at sufficiently large $U$ is preempted by a Mott transition as $T$ is reduced. This is evidenced through the insulating behavior of $-\mathrm{Im}[\Sigma(i\omega_n)]$ at low $T$, as shown in Fig. 3.



\end{document}